\documentclass[%
reprint,
superscriptaddress,
amsmath,amssymb,
aps,
]{revtex4-2}
\usepackage{algpseudocode}
\usepackage[ruled]{algorithm}

\renewcommand{\algorithmicrequire}{\textbf{Input:}}  
\renewcommand{\algorithmicensure}{\textbf{Output:}} 
\usepackage{graphicx}
\usepackage{float}
\usepackage{epstopdf}
\usepackage{stfloats}

\usepackage{bm}
\usepackage{ragged2e}

\begin{document}
\preprint{APS/123-QED}

\title{Performance of the quantum MaxEnt estimation in the presence of physical symmetries}

\author{Diego Tielas}
    \affiliation{Departamento de F\'isica, IFLP - CONICET, Universidad Nacional de La Plata, C.C. 67, 1900 La Plata, Argentina}
    \affiliation{Departamento de Ciencias B\'asicas, Facultad de Ingenier\'ia, Universidad Nacional de La Plata, La Plata, Argentina}

\author{Marcelo Losada}%
\affiliation{%
 Universidad Nacional de Córdoba, FAMAF, CONICET, Córdoba, Argentina
}%
    
\author{Lorena Reb\'on}
 \email{rebon@fisica.unlp.edu.ar}
    \affiliation{Departamento de F\'isica, IFLP - CONICET, Universidad Nacional de La Plata, C.C. 67, 1900 La Plata, Argentina}
    \affiliation{Departamento de Ciencias B\'asicas, Facultad de Ingenier\'ia, Universidad Nacional de La Plata, La Plata, Argentina}

\author{Federico Holik}
\affiliation{Departamento de F\'isica, IFLP - CONICET, Universidad Nacional de La Plata, C.C. 67, 1900 La Plata, Argentina}



\begin{abstract}

When an informationally complete measurement is not available, the reconstruction of the
density operator that describes the state of a quantum system can be accomplish, in a reliable way, by adopting the maximum entropy principle (MaxEnt
principle), as an additional criterion, to obtain the least biased estimation.
In this paper, we study the performance of the MaxEnt method for quantum state estimation when there is prior information about symmetries of the unknown state. We explicitly describe how to work with this method in the most general case, and present an algorithm that allows to improve the estimation of quantum states with arbitrary symmetries.
Furthermore, we implement this algorithm to carry out numerical simulations estimating the density matrix of several three-qubit states of particular interest for quantum information tasks. We observed that, for most states, our approach allows to considerably reduce the number of independent measurements needed to obtain a sufficiently high fidelity in the reconstruction of the density matrix. Moreover, 
we analyze the performance of the method in realistic scenarios, showing that it is robust
even when considering the effect of finite statistics, and under the presence of typical experimental noise.  
\end{abstract}

\maketitle

\section{Introduction}

A fundamental problem that crosses the interdisciplinary field of quantum information science is estimating the unknown state of a quantum system \cite{paris2004quantum}. This task can be achieved by performing a quantum state tomography (QST), which
consists of obtaining statistical information about the state
by measuring many copies of the system in different basis. Ideally,
when a set of informationally complete measurements is available, the density matrix of the initially unknown state
can be univocally determined within experimental errors. However, the number of independent measurements needed to perform a full QST scales exponentially with the number of particles in the system,
becoming impractical in many circumstances of interest.
In this context, quantum 
state estimation techniques with partial information become of great relevance~\cite{Flammia_2012,Rebon17,zambrano2020}.

The general inference method based on the maximum entropy principle (MaxEnt
principle) is considered as
a powerful method for estimating probability distributions in a wide range of
probabilistic models~\cite{Canosa-Plastino-Rossignoli-PRA-MaxEnt,Plastino-Portesi-PRA-MaxEnt,Diambra-Plastino-PRE-MaxEnt-1995,Rebollo-Neira-Plastino-PRE-MaxEnt-2001,Sinatra-MaxEntRandomWalks-2011,Goswami-Prasad-PRD-MaxEnt-2013,Tkacic-MaxEntNeuralNetwork-2013,DEMARTINO2018,Tang2021}.
This principle asserts that the most suitable probability
distribution compatible with a given set of empirical data (or constraints) is the one with the largest
entropy \cite{Jaynes-1957a,Jaynes-1957b}. Moreover, 
by using the von
Neumann entropy, which is considered to be the quantum extension of the Shannon entropy,
it has been proven 
to be useful for
a reasonable estimation of a quantum state
from incomplete data
\cite{MaxEnt-Tomography,Teo-2011,Goncalves-2013-MaxEntTomography,MaxEnt-PRXQuantum-2021}. 

As expected, the performance of the MaxEnt method in the quantum domain depends on the number of independent measurements performed. In fact, the fidelity -or any other measure of similarity- between the obtained state and the one that is being estimated drops off as the number of independent measurements decreases \cite{Goncalves-2013-MaxEntTomography}. However, the loss of fidelity could be compensated in those cases in which there is some \textit{a priori} information available about the system. For instance, the problem of determining the MaxEnt state for generalized probabilistic models under 
symmetry constraints was studied in Ref.~\cite{Holik-Massri-Plastino-IJGMMP-2016}, and the specific set of equations that solves the problem for the finite dimensional quantum case was recently developed in Ref.~\cite{MaxEntSolutions}. This covers many situations of interest in which the experimenter has prior information about a particular symmetry of the state to be determined. As examples, we can mention systems with a known symmetric Hamiltonian and quantum protocols involving symmetric states \cite{Pollatesek-2004,Ouyang-2014,Ouyang-2021}. In particular, 
many applications to quantum information tasks
with spin systems~\cite{Lehmann2009,morton2018storing} or photons~\cite{Joshi_2021,Boson_Sampling} involve states which are permutationally invariant.

In this paper, we study the performance of the MaxEnt principle for estimating quantum states which are known to possess symmetries. Tomographic methods adapted to \textit{particular} symmetries were studied in previous works. For example, the case of permutationally invariant states was addressed in \cite{PermutationallyInvariantQT,Moroder_2012}. Such methods rely on the particular features of the symmetries involved and require a data set that is informationally complete for the given class of states. 
Using as a starting point the formalism elaborated in \cite{MaxEntSolutions}, here we put the focus in developing a reconstruction algorithm that works for \textit{arbitrary} symmetries and in situations where the available information is incomplete for performing a full tomography. We show how to apply it, also assessing its performance in different examples of interest in quantum information theory. Unlike other methods, our algorithm is general and only needs, as extra inputs, the generators of the group that represent the given symmetry (for symmetries based on finite groups, as in the case of permutation symmetry), or the generators of the Lie algebra associated to the group (if the symmetry is based on connected Lie groups), without any other dependence on the mathematical features of the particular symmetry. This is an 
important point, since the mathematical treatment of symmetries can be cumbersome in many cases, requiring advanced knowledge of group theoretical techniques and linear algebra. On the contrary, our method allows to deal with arbitrary symmetries in a simple and systematic way, by only identifying its generators,
something that, in most cases, is equivalent to the specification of the symmetry itself. As a main result, we show that the way in which prior knowledge about symmetries is incorporated into the MaxEnt estimation, gives a substantial reduction in the number of independent measurements needed for a reasonable reconstruction of the target state. 

The paper is organized as follows: In Section~\ref{formalism}, we describe how to include the information about the symmetry of the state to be estimated. In Section~\ref{s:GeneralAlgorithm}, the MaxEnt method with symmetries constraints is presented in an algorithmic way, so that it can be easily applied in practical situations and combined with other existing estimation techniques. Its performance is analyzed in Section~\ref{s:NumericalSimulations}, by numerical simulations of the reconstruction process of different class of three-qubit states. We start by characterizing the performance of the standard quantum MaxEnt technique on a sample of arbitrary states randomly chosen.
This is then compared to its performance on samples of states with a particular symmetry, for which thousands of permutationally invariant states and Werner states, randomly chosen, are used as statistical samples. We highlight that, in average, the number of measurements needed for a reasonable estimation of the state grows considerably in the latest cases, i.e., the performance of standard MaxEnt drops off when the source of quantum states is biased.
Next, we focus on analyzing how the estimation of these states is improved with the use of our algorithm based on the MaxEnt technique with symmetries. As symmetric states of particular interest for quantum information, the estimation of three-qubit cat-like states and Dicke states, is also studied in detail. In addition, we analyze the performance of the method in the presence of simulated experimental noise and finite statistic, compatible with photonic implementations.  
Finally, after discussing the main results and open problems, we draw our conclusions in Section~\ref{conclusions}.

\section{Formalism}\label{formalism}

The MaxEnt principle in the quantum scenario postulates that the most
suitable state $\rho^{\mathrm{ME}}$ compatible with the 
available data is the one with largest von Neumann entropy.
In the standard formulation of the quantum MaxEnt problem \cite{MaxEnt-Tomography}, the 
only information about the target state $\rho$ is given in terms of the expectation values $\{a_i\}$ of a set of observables. The maximization problem involves $r$
Hermitian operators $A_i$ ($1\leq i \leq r$) which correspond to the 
$r$ observables to be measured, and thus the constraints associated with such observables are
\begin{eqnarray}\label{e:Conditions}
	\langle A_{i}\rangle&=&  \mbox{Tr} \left(A_i\rho^{\mathrm{ME}} \right)  = a_i, ~~~ \forall \,  i= 1, \ldots, r.
\end{eqnarray}
The MaxEnt solution, i.e., the state that satisfies the constraints and maximizes the von Neumann entropy, is given by
\begin{equation}
\rho^{\mathrm{ME}} = \frac{e^ {\sum_{i=1}^r \lambda_{i}A_{i}}}{Z},
\end{equation}
where $Z = \mbox{Tr}(e^{\sum_{i=1}^r \lambda_{i} A_{i} })$, and the Lagrange multipliers, $\{\lambda_i\}_{i= 1}^r$, are given by the relations
\begin{equation}\label{e:ConditionsA}
a_{i}= \frac{\partial}{\partial\lambda_{i}}\ln Z,\quad  1\leq i \leq r.
\end{equation}

In this work, we will consider some prior information given in terms of symmetries of the state. When this is the case, the maximization problem also includes a group $\mathcal{G}$ representing such symmetries. In this scenario, the symmetric MaxEnt estimator $(\rho^{\mathrm{SME}})$ must be invariant under the action of the symmetry group, that is 
\begin{eqnarray}
U_{g} \rho^{\mathrm{SME}} U_{g}^{\dag}= \rho^{\mathrm{SME}},
\end{eqnarray} 
for all $g \in \mathcal{G}$, and $U_{g}$ the unitary operator representing $g$.
For quantum systems described by finite dimensional Hilbert spaces, 
this problem can be reformulated as that of finding the density matrix
$\rho^{\mathrm{SME}}$ which maximizes the von Neumann entropy and, besides the conditions of Eq.~(\ref{e:Conditions}),
satisfies the following constraints \cite{MaxEntSolutions}
\begin{eqnarray}\label{e:ConditionsB}
	\langle [i Q_k,O_j] \rangle&=&
	\mbox{Tr}\left(i  [Q_k,O_j]\rho^{\mathrm{SME}}\right)=0,\nonumber\\ ~~\forall k \in I,
	&~& \forall \,  j= 1, \ldots, m^2,
\end{eqnarray}
where $\{O_j\}_{1\leq j \leq m^2}$ is a basis of the space of Hermitian operators associated to the Hilbert space $\mathcal{H}$ (with $\dim(\mathcal{H})=m$), $\{Q_k\}_{k\in I}$ are the generators of the Lie algebra $\mathcal{L}(\mathcal{G})$ of the group $\mathcal{G}$, and $I$ is a set of indexes whose cardinal equals the dimension of $\mathcal{L}(\mathcal{G})$. 
Since $i[Q_k,O_j]$, is also an Hermitian operator, it is associated with a new observable that will be called \textit{auxiliary observable}. 
Hence, the symmetry conditions are reformulated in terms of, in principle, $\dim(\mathcal{L}(\mathcal{G})) \times m^2$ \textit{extra} mean
values constraints equal to zero, and the solution has the same form as in the standard quantum MaxEnt problem. Explicitly, it is given by
\begin{equation}\label{solution_LD}
	\rho^{\mathrm{SME}} = \frac{e^ {\left(\sum_{i=1}^r
			\lambda_{i}A_{i}+\sum_{k\in
			I}\sum_{j=1}^{m^2}\gamma_{k,j}[iQ_k,O_j]\right)}}{Z},
\end{equation}
\noindent where $Z=
\mbox{Tr}\left(e^{\left(\sum_{i=1}^r\lambda_{i}A_{i}+\sum_{k\in I}\sum_{j=1}^{m^2}\gamma_{k,j}[iQ_k,O_j]\right)}\right)$, and the
Lagrange multipliers, $\{\lambda_i\}_{i= 1}^r$ and $\{\gamma_{k,j}\}_{k \in I}^{j = 1, \ldots, m^2}$, which satisfy the relations
\begin{align}
	a_{i}&= \frac{\partial}{\partial\lambda_{i}}\ln Z,\quad  1\leq i \leq r, \nonumber\\
	0 &= \frac{\partial}{\partial\gamma_{k,j}}\ln Z,\quad  1\leq j \leq
	m^2, \, k \in I. \label{lagrangian multiplier 2}
\end{align}

It is important to mention that, in general, the auxiliary observables $\{i[Q_{k},O_{j}]\}_{k\in I}^{j=1,\dots, m^2}$ will not be linearly independent. In such a case, the largest linearly independent set $\{ i[Q_{k},O_{j}]\equiv \tilde{A}_{i}\}_{1\leq i\leq R}$ should be determined before computing $\rho^{\mathrm{SME}}$, which can be finally expressed as
\begin{eqnarray}\label{general solution}
	\rho^{\mathrm{SME}} &=& \frac{e^ {\left(\sum_{i=1}^r
			\lambda_{i}A_{i}+\sum_{i=1}^R
			\tilde{\lambda}_{i}\tilde{A}_{i}\right)}}{\tilde{Z}},\nonumber\\
\tilde{Z}&=&\mbox{Tr}\left(e^{\left(\sum_{i=1}^r
			\lambda_{i}A_{i}+\sum_{i=1}^R
			\tilde{\lambda}_{i}\tilde{A}_{i}\right)}\right)\\
			a_{i}&=& \frac{\partial}{\partial\lambda_{i}}\ln \tilde{Z},\quad  1\leq i \leq r, \nonumber\\
	0 &=& \frac{\partial}{\partial\tilde{\lambda}_{i}}\ln \tilde{Z},\quad  1\leq i \leq R.\nonumber 
\end{eqnarray}

In many situations, the set of symmetry constraints, given by Eqs.~(\ref{e:ConditionsB}), significantly reduces the dimensionality of the problem. These equations restrict the search of the MaxEnt state to a lower dimensional space within the set of density matrices $\mathcal{M}$. In the general quantum MaxEnt problem, the number of independent observables to be measured in order to obtain a reasonable estimation is less (or equal in the worst case) than the dimension of $\mathcal{M}$. Including prior information about symmetries allows to take advantage of relevant information, without the need to resort to extra measurements. This strategy could significantly reduce the amount of experimental resources required to obtain the same level of accuracy in the estimation.

\section{An algorithm for state estimation with symmetry constraints}\label{s:GeneralAlgorithm}  
In this section we present a concrete algorithm for implementing the formalism displayed 
in the previous section. We provide a recipe for getting advantage of the prior information encoded in terms of symmetries of the state to be estimated. For a practical implementation the steps to follow are:
\begin{itemize}
    \item Identify the symmetries of the state to be estimated. This step involves previous knowledge of certain characteristics of the state or about the process that generates it.
    \item If the symmetries of the state to be estimated can be expressed using a \textit {continuous group} $\mathcal{G}$, identify a basis $\{Q_{1},Q_{2},\ldots,Q_{s}\}$ of the Lie algebra $\mathcal{L(G)}$. For a \textit {discrete group}, provide a list of all the operators representing the action of the group generators.
    \item Choose a basis $\{O_{j}\}_{1\leq j \leq m^2}$ of the space of Hermitian operators acting on the Hilbert space $\mathcal{H}$.
    \item Compute the set of operators $\{i[Q_{k},O_{j}]\}_{k\in I}^{j=1,\dots, m^2}$, representing the auxiliary observables. From this set, extract the largest set of linear independent operators  $\{\tilde{A}_{i}\}_{1\leq i\leq R}$.
    \item Choose the set of observables $\{A_{i}\}_{1\leq i\leq r}$ to be measured. To add significant information, they have 
    to be linearly independent with regard to the auxiliary observables $\{\tilde{A}_{i}\}_{1\leq i\leq R}$.
    \item Find the MaxEnt estimator using as inputs the data obtained from the measured observables $\{a_{i}\}_{1\leq i \leq r}$. The expectation values of the auxiliary observables must be set to zero.  
\end{itemize}
In Algorithm \ref{alg:Framwork} we show the pseudo-code to obtain the MaxEnt density matrix. It was implemented based on gradient descent optimization algorithm.

\begin{algorithm}[H]
\caption{MaxEnt with (or without) symmetries}\label{alg:Framwork}  
\begin{algorithmic}[1]
\Statex
\Require observables $\{A_{i}\}_{1\leq i\leq r}$; mean values $\{a_i\}_{1\leq i\leq r}$; \;\;\;\;\;generators $\{Q_{1},Q_{2},\ldots,Q_{s}\}$ of $\mathcal{L(G)}$.
\Ensure MaxEnt density matrix $\rho_r^{\mathrm{SME}} ~ (\rho_r^{\mathrm{ME}})$.
\Statex
\If {\textit{symmetries are known}}
  \State Extract the maximal subset of LI auxiliary observables $\{\tilde{A}_{i}\}_{1\leq i\leq R}$;
  \Statex
 \State Add a maximal subset of observables $\{A_i\}_{1\leq i\leq r}$ so that the resulting set is LI;  
\State $\displaystyle \min_{\bar{\lambda}}\lbrace \sum^{r}_{i=1} \left(\mathrm{Tr}\left(A_i\rho_r^{\mathrm{SME}}(\bar{\lambda}) \right)-a_i\right)^2$
\Statex $\;\;\;\;\;\;\;\;\;\; +\displaystyle \sum^{R}_{i=1} \mathrm{Tr}\left(\tilde{A}_i \rho_r^{\mathrm{SME}}(\bar{\lambda})\right)^2\rbrace.$
\Statex
\Else 
\State Extract the maximal subset of LI observables $\{A_i\}_{1\leq i\leq r}$;
\State $\displaystyle \min_{\bar{\lambda}} \sum^{r}_{i=1} \left(\mathrm{Tr}\left(A_i\rho_r^{\mathrm{ME}}(\bar{\lambda}) \right)-a_i\right)^2$.
\Statex
\EndIf
\Statex
\end{algorithmic}
\end{algorithm}


\section{Characterizing the performance of the MaxEnt method}\label{s:NumericalSimulations}

In order to investigate 
the performance of the MaxEnt method when symmetry constraints are considered, we have carried out numerical simulations for the reconstruction of three-qubit states~\footnote{
The empirical results used in this work were obtained via synthetic experiments. The algorithms were implemented as MATLAB and Phyton codes. They may be obtained from the authors upon reasonable request.}. We have considered two types of states, both of relevance in quantum information: permutationally invariant states and Werner states. For comparison reasons, we start by analyzing the standard MaxEnt estimation,
i.e., the MaxEnt estimation method without symmetry constraints.

\subsection*{MaxEnt method without symmetry constraints}
Let us consider an informationally complete set of observables $\{A_{i}\}_{1\leq i\leq q}$ and a given target state $\rho$. Then, we can compute the corresponding expectation values $a^{\rho}_{i}=\mathrm{Tr}(\rho A_{i})$, with $1\leq i\leq q$. To analyze the dependence of the MaxEnt estimator on the amount of available data, we have considered a sequence $\{S_{r}\}_{1\leq r\leq q}$ of sets of expectation values, defined as
\begin{eqnarray}
\begin{cases}
 S_{1}=\{a^{\rho}_{1}\}  \\ 
 S_{r}=S_{r-1}\cup\{a^{\rho}_{r}\} \text{ if } 2\leq r\leq q. 
\end{cases}
\end{eqnarray}
That is, each set in the sequence contains all the expectation values in the previous set, and one more, not contained in the previous set. Intuitively, as the number $r$ grows, the quality of the estimate $\rho_{r}^{\mathrm{ME}}$ should improve. Indeed, for $r=q$, it is expected that $\rho_{r}^{\mathrm{ME}}=\rho$.

Representative examples of the performance of the standard MaxEnt estimation are shown in Fig.~\ref{f:1000MaxEntSolo}. Algorithm \ref{alg:Framwork} (without symmetries) was used to estimate a set of 1000 three-qubit pure states, randomly chosen and uniformly distributed throughout the Hilbert space, according to the Haar measure. As informationally complete set of observables we have used both the Pauli tensor-product operators (Figs.~\ref{f:1000MaxEntSolo}(a) and \ref{f:1000MaxEntSolo}(b)) and a SIC-POVM (Fig.~\ref{f:1000MaxEntSolo}(c)), in order to also assess the dependence of the estimation on the chosen set of measurement basis. 
To quantify the quality of each estimation we have computed the fidelity
$F \equiv \mathrm{Tr}\left(\sqrt{\sqrt{\rho}\rho_{r}^{{\mathrm{ME}}}\sqrt{\rho}}\right)$ between the target state and the estimated state  \cite{Jozsa1994}, which is essentially a measure of the geometrical proximity in the Hilbert space. As a criterion of a good estimation, it is desirable a fidelity value close to one. Each point of the plots shows the fidelity of reconstruction obtained from the expectation values $S_r$, of a subset of $r$ Pauli tensor-product or SIC-POVM operators.  
For this random sample of states, we see that it only takes between 25 and 35, out of a total of 63, to reach an average  fidelity value $\bar{F}$ greater than 0.95. 
In each case, we also show the standard deviation of the fidelity 
to account for how much the performance of the MaxEnt estimation will depend on the particular state to be inferred.
In accordance with what is shown in Ref.~\cite{Goncalves-2013-MaxEntTomography}, our results indicate a good overall performance of the standard MaxEnt method when is applied to a sample of
pure states with an uniform probability distribution (unbiased source).
%
\begin{widetext}
\begin{minipage}{\linewidth}
\begin{figure}[H]
\centering
\includegraphics[width=1\textwidth]{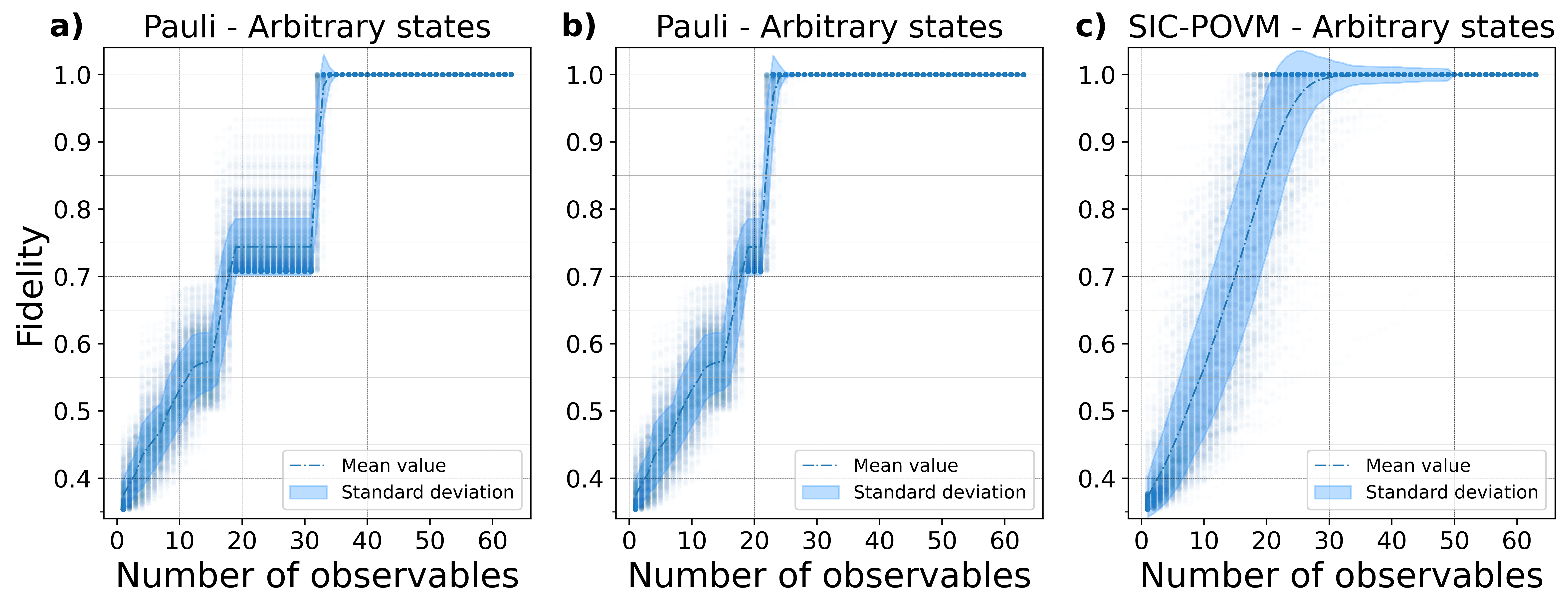}
\vspace{-0.5cm}
\caption{\label{f:1000MaxEntSolo}\small{Performance of the standard quantum MaxEnt estimation for three-qubit states.
Each point in the figures represents the fidelity of reconstruction $F$ between a target state and the one obtained from an expectation value set $S_r$, corresponding to a subset of $r$ Pauli \textbf{(a)}-\textbf{(b)}, or SIC-POVM \textbf{(c)} operators. We have run the algorithm for 1000 states randomly chosen according to the Haar measure in the entire Hilbert space. The dot-dashed lines represent the mean value of $F$ while the shaded area indicates its standard deviation. All figures were built from the same sample of target states. From panel \textbf{(a)} to panel \textbf{(b)} the ordering of the Pauli operators was changed.
}}
\end{figure}    
\end{minipage}
\end{widetext}

It should be noted that, for $r < q$, the performance of the estimation may depends on the different possibilities of selecting a set of $r$ observables.
Indeed, when the estimation is based on Pauli tensor-product operators, there is a particular subset  
of these observables, which do not add substantial information for most of the  
sampled states. This can be inferred from the plateau observed in the mean fidelity curve in Fig.~\ref{f:1000MaxEntSolo}(a), and the fact that the sequence $\{S_{r}\}_{1\leq r\leq q}$ used in the estimation was the same for all states. 
After identifying that subset of observables (from observable 21 to 31 of the list described in \footnote{In Fig.~\ref{f:1000MaxEntSolo}(a), the ordering of the observables from which the sequence  $\{S_{r}\}_{1\leq r\leq q}$ was built, as follows.
Let $\Omega=\{I_2,\sigma_{x},\sigma_{y},\sigma_{z}\}$ be the set of Pauli operators together the identity single-qubit operator, $I_2$. Then we have obtained all possible tensor products of single-qubit operators, $\Omega\otimes\Omega\otimes\Omega$, in this order:
$I_2\otimes I_2\otimes I_2, I_2\otimes I_2\otimes\sigma_x, I_2\otimes I_2\otimes\sigma_y, I_2\otimes I_2\otimes\sigma_z, I_2\otimes \sigma_x\otimes I_2, I_2\otimes \sigma_x\otimes \sigma_x, I_2\otimes\sigma_x\otimes \sigma_{y}, I_2\otimes\sigma_x\otimes \sigma_{z},I_2\otimes\sigma_y\otimes I_2,I_2\otimes\sigma_y\otimes \sigma_{x},\dots,I_2\otimes\sigma_z\otimes I_2,I_2\otimes\sigma_z\otimes \sigma_{x},\dots,\sigma_z\otimes I_2\otimes I_2,\sigma_z\otimes I_2\otimes \sigma_{x},\dots$. For the sake of completeness, we have included here the identity tensor-product operator $I_2\otimes I_2\otimes I_2$, but it was not taken into account in the simulations. In a real experiment, this is measured with the only purpose of obtaining the normalization with regard to the total number of counts.}), 
we performed the reconstruction of the same sample of states, but now using a different sequence $\{\tilde{S}_{r}\}_{1\leq r\leq q}$, so that the average values of these observables appear at the end of the sequence. The results obtained for the new ordering are displayed in Fig.~\ref{f:1000MaxEntSolo}(b). The number of observables needed to reach a mean fidelity greater than 0.95 has now been reduced from 35 to 25.
However, when the estimation is based on SIC-POVMs, a smooth and ever-increasing behavior of the mean fidelity is observed (Fig.~\ref{f:1000MaxEntSolo}(c)). It is reasonable to think that this behavior is linked to the geometric structure of a SIC-POVM, whose projections are oriented in equiangular directions of the generalized Bloch sphere. Such a measurement minimizes the informational overlap or redundancy, in the sense that the information extracted from each single measurement is maximum, and thus, they are less biased than Pauli measurements. Hence, as the states to be estimated are uniformly distributed in all possible directions of the Hilbert space, the reconstruction performs better when measurements are isotropically oriented as in the case with SIC-POVMs.

But what happens if our source of quantum states has a bias, so the states generated (and to be estimated) are not uniformly distributed in the Hilbert space?
On the one hand, often the source will produce states with well-defined properties inherent to the experimental realization, which might be very far from those uniformly generated. On the other hand, many tasks or algorithms in quantum information processing require or (work optimally) based on states that have a specific symmetry, or result in such states.   
Then, for an experimenter dealing with these kind of sources, the standard MaxEnt method might be not so useful (or at least not optimum) in terms of efficiency. 

In the following subsections, we will illustrate, through several examples, the importance of a detailed study of the MaxEnt method in different scenarios. It will show, quantitatively, how its performance drops off drastically when the states to be estimated have a certain symmetry, but this prior information is not taken into account in the estimation process.
At the same time, we show how the estimation can be improved significantly when the information about the symmetries of the state is incorporated, by appealing to the algorithm described in Section \ref{s:GeneralAlgorithm}. Before discussing the results, we will explicitly describe how to obtain the generators $\{Q_{1},Q_{2},\ldots,Q_{s}\}$ and the set of linear independent auxiliary observables $\{\tilde{A}_i\}_{1\leq i \leq,R}$, for the examples to be considered here. 

\subsection*{MaxEnt method with symmetry constraints}

\subsubsection{Permutationally invariant states}

To fix ideas, we start by considering a source that generates \textit{permutationally invariant states}. For pure states, these are symmetric or antisymmetric under the exchange of any two  particles of the system. For example, given the tensor product state $|\psi\rangle=|\psi_{1}\rangle\otimes\ldots\otimes|\psi_{i}\rangle\otimes\ldots\otimes|\psi_{j}\rangle\otimes\ldots\otimes|\psi_{N}\rangle$, for any pair $i,j=1,\ldots,N$, the action of the permutation operator 
$P_{ij}$ ($i\neq j$) is defined by
\begin{equation}\label{permutator_operattor}
    P_{ij}|\psi\rangle=|\psi_{1}\rangle\otimes\ldots\otimes|\psi_{j}\rangle\otimes\ldots\otimes|\psi_{i}\rangle\otimes\ldots\otimes|\psi_{N}\rangle,
\end{equation}
Then, for any pair $(i,j)$, a permutationally invariant state $|\psi\rangle$ must satisfy the relation $P_{ij}|\psi\rangle=\pm|\psi\rangle$, where the ``$+$" stands for bosons and the ``$-$" for fermions. In the most general case, a quantum system having permutational symmetry is one whose state is described by a density operator $\rho_{\mathrm{(PI)}}$ that satisfies the relation
\begin{eqnarray}
P_{ij}\rho_{\mathrm{(PI)}} P_{ij}=\rho_{\mathrm{(PI)}}, ~~~\forall~ i,j=1,\dots, N.
\end{eqnarray}
Any other permutation of $n\leq N$ particles can be written as a product
of $P_{ij}$'s and therefore $\rho_{\mathrm{(PI)}}$ is invariant under a general $n$-particle permutation operation.
These operations form a \textit{discrete group}, which has a finite set of generators. For example, for a composite system of $N$ identical particles there are $N-1$ generators, 
and we can consider as a set of generators $\{Q_k\}_{1\leq k \leq N-1}$, the one given by $\{P_{12},P_{13},P_{14},\ldots,P_{1N}\}$. In the computational basis, the matrix representation of the elements in $\{P_{1,k+1}\}_{1\leq k \leq N-1}$ can be computed directly. Then, 
the auxiliary observables given in Eq.~(\ref{e:ConditionsB}) must be calculated as:
\begin{eqnarray}\label{Aux_Perm}
\{[iP_{1,k+1},O_j]\}_{k=1,\ldots,N-1}^{j=1,\ldots,2^N\times 2^N}.
\end{eqnarray}

Not all the elements of the set in Eq.~(\ref{Aux_Perm}) 
are necessarily linearly independent. 
In fact, for three qubits there are $44$ linearly independent auxiliary observables that form the set $\{\tilde{A}_i\}_{1\leq i \leq,R}$ needed to run the MaxEnt (with symmetries) algorithm described in Algorithm~\ref{alg:Framwork}.

\subsubsection{Werner states}

We analyze here how to proceed with the family of \textit{Werner states}~\cite{Werner1989,Eggeling2001}. In the case of a system composed of $N$ qubits, these states can be defined as those that are invariant under the action of the group 
\begin{equation}\label{werner_group}
\mathcal{G}_{N}=\{\otimes^{N} U\,\,\mid\,\,U\in U(2)\}, 
\end{equation}
that is, a Werner state satisfies the relation 
\begin{eqnarray}\label{sym_Werner}
(\otimes^{N} U)\rho_{\mathrm{(W)}}(\otimes^{N}U)^{\dagger}=\rho_{\mathrm{(W)}}, 
\end{eqnarray}
for all unitary operators $U$ acting on the single-qubit space. Thus, these states have the symmetry defined by the action of the \textit{continuous group} $\mathcal{G}_{N}$.

In order to find the generators of its Lie algebra $\mathcal{L}(\mathcal{G}_N)$, we will first analyze the case of two qubits. Let $I_2$ be the $2\times 2$ identity matrix. Since $ia\otimes I_2$ commutes with $I_2\otimes ia$, for all $ia\in \mathcal{L}(U(2))$, it is verified that 
\begin{eqnarray}
e^{i(a\otimes I_{2}+I_{2}\otimes a)}&=&e^{ia\otimes I_{2}}e^{I_{2}\otimes ia}=
(e^{ia}\otimes e^{I_{2}})(e^{I_{2}}\otimes e^{ia})\nonumber\\&=&e^{ia}\otimes e^{ia}= U\otimes U,
\end{eqnarray}
which shows that all the elements of $\mathcal{L}(\mathcal{G}_{2})$ are of the form $ia\otimes I_{2}+I_{2}\otimes ia$, with $ia\in \mathcal{L}(U(2))$. Similarly, the expression 
$ia\otimes I_{2}\otimes I_{2}\otimes\ldots \otimes I_{2}+I_{2}\otimes ia\otimes I_{2}\otimes\ldots\otimes I_{2} +I_{2}\otimes I_{2}\otimes ia \otimes I_{2}\otimes\ldots\otimes I_{2}+\ldots + I_{2}\otimes\ldots \otimes I_{2}\otimes ia$, where each of the $N$ terms in the sum has exactly $N$ tensor products,
can be written as $\bigotimes_{\ell=1}^{N}e^{ia}=\bigotimes_{\ell=1}^{N}U$.
Hence, any element of $\mathcal{L}(\mathcal{G}_{N})$ belongs to the set $\{\sum_{\ell=1}^{N}ia^{(\ell)}\otimes I_2\,\,|\,\,ia\in \mathcal{L}(U(2))\}$, given the compact notation $ia^{(\ell)}\otimes I_2$, which means that $ia$ is placed in the the $\ell$-th position of the tensor product, and there is an identity matrix in each of the other positions. 
As the set of matrices   $\{I_2,\sigma_x,\sigma_y,\sigma_z\}$ (i.e., the three Pauli matrices together with $I_2$, and which we will refer to as $\{\sigma_0,\sigma_1,\sigma_2,\sigma_3\}$) 
forms a basis of $\mathcal{L}(U(2))$, we can decompose $ia$ in terms of this set, obtaining
\begin{eqnarray}
\sum_{\ell}^{N}ia^{(\ell)}\otimes I_2=\sum_{k=0}^{3}\alpha_{k}(\sum_{\ell=1}^{N}i\sigma_{k}^{(\ell)}\otimes \sigma_0),
\end{eqnarray}
which shows that the generators $\{Q_k\}_{1\leq k \leq 4}$ of $\mathcal{L}(\mathcal{G}_N)$ are given by the set $\{\sum_{\ell=1}^{N}\sigma_{k-1}^{(\ell)}\otimes \sigma_0\}_{1\leq k \leq 4}$. Finally, and after choosing a basis $\{O_{j}\}_{1\leq j \leq 2^{N}\times 2^{N}}$ of Hermitian operators acting on $\bigotimes^{N}\mathbb{C}^{2}$,
the auxiliary observables can be computed as
\begin{eqnarray}
\{[i\sum_{\ell=1}^{N}\sigma_{k-1}^{(\ell)}\otimes \sigma_0,O_{j}]\}_{k=1,\dots,4}^{j=1,\dots,2^{N}\times 2^{N}},
\end{eqnarray}
 from where the largest set of linear independent observables  $\{\tilde{A}_{i}\}_{1\leq i\leq R}$ should be extracted.

\subsection*{Numerical simulations and analysis results}
In order to test their performance, we ran the algorithms for the standard MaxEnt and MaxEnt with symmetry constraints to estimate states with two different types of symmetries, namely, permutationally invariant and Werner states. For each family, a set of 1000 states was randomly chosen. For the sake of comparison, the results of the estimation with and without taking into account the symmetries of the states are displayed in Fig.~\ref{f:1000QOffOffSIC}.
%
\begin{widetext}
\begin{minipage}{\linewidth}
\begin{figure}[H]
\centering
\includegraphics[width=1\textwidth]{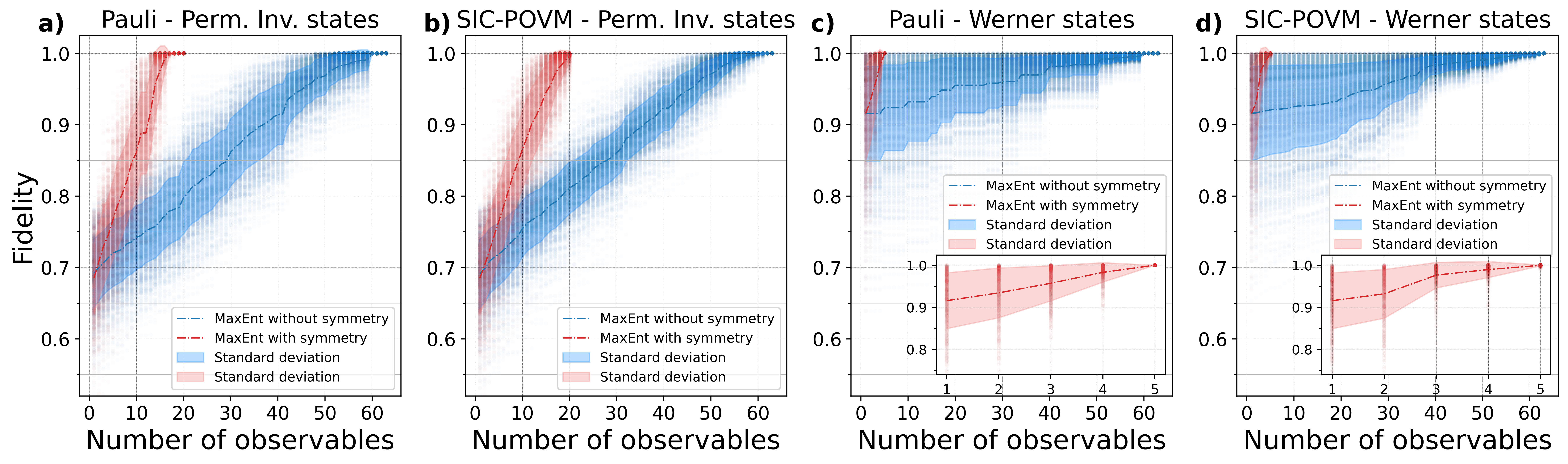}
\vspace{-0.5cm}
\caption{\label{f:1000QOffOffSIC}\small{Performance of the quantum MaxEnt estimation for three-qubit states by following the standard algorithm (\textit{blue}) and the algorithm that includes the knowledge of symmetries of the state (\textit{red}). Each point in these figures represents the fidelity of reconstruction between a permutationally invariant state, in panel \textbf{(a)} and panel \textbf{(b)}, or a Werner state in panel \textbf{(c)} and panel \textbf{(d)}, and the one obtained from an expectation value set $S_r$, corresponding to a subset of $r$ Pauli (panel \textbf{(a)} and panel \textbf{(c)}) or SIC-POVM (panel \textbf{(b)} and panel \textbf{(d)}) operators. We have run the algorithm for 1000 states randomly chosen. The dot-dashed lines represent the mean value of the fidelity while the shaded area indicates its standard deviation. All figures were built from the same sample of target states.}}
\end{figure}    
    \end{minipage}
\end{widetext}

Figures~\ref{f:1000QOffOffSIC}(a) and \ref{f:1000QOffOffSIC}(b) show the fidelity of reconstruction of permutationally invariant states obtained from the expectation values $S_r$, of a set of $r$ Pauli tensor-product operators or SIC-POVM operators, respectively.  
It can be seen that even if for some states the performance of the standard quantum MaxEnt method (\textit{blue points}) is still good, globally, it clearly drops off with respect to the previous case (Figs.~\ref{f:1000MaxEntSolo}(a)-\ref{f:1000MaxEntSolo}(c)) where the states to be estimated were randomly chosen from an homogeneous sample (unbiased source). 
In the present case, between 40 and 50 observables are needed to reach a mean fidelity value greater than 0.95, which represents about 15 extra observables to be measured compared to the previous case. We can also observe a decrease in the performance of the standard method when the estimation is performed on a sample of Werner states (Figs.~\ref{f:1000QOffOffSIC}(c) and \ref{f:1000QOffOffSIC}(d)). In this case, although the mean fidelity reaches a value above 0.90 for a few observables, the growth of the curve and its convergence to the optimal value are slow. 

In all the previous examples we observe that, in comparison with an unbiased source, more observables must be considered in the standard MaxEnt estimation to achieve, in average, an acceptable fidelity value in the reconstruction of an unknown state that posses a certain symmetry (biased source). However, when symmetry constraints are incorporated to the MaxEnt method (\textit{red points}), a different behavior of the fidelity as a function of the available set of expectation values $S_r$, is obtained.
For permutationally invariant states (Figs.~\ref{f:1000QOffOffSIC}(a) and \ref{f:1000QOffOffSIC}(b)), 
the information obtained from about 15 observables (out of a total of 63), is enough to reach a mean fidelity value of 0.95, 
while for Werner states (Figs.~\ref{f:1000QOffOffSIC}(c) and \ref{f:1000QOffOffSIC}(d)),
the same mean fidelity value is reached by considering only 3 observables. 

The above results clearly show that the variant of the MaxEnt method studied here, allows to incorporate the prior knowledge about the symmetries of the state, so that, it can be used effectively for an optimal reconstruction (in terms of reduction of independent measurements) of its density matrix. 
As expected, such reduction in the number of independent measurements required to reach a given accuracy in the estimation, will depend on the particular symmetry.
For example, by looking at the Hermitian elements of the symmetric tensor-algebra of order $N$, which is generated by $d\times d$ complex matrices (see Eq.~(\ref{permutator_operattor})), we find that the number of independent real parameters needed to \textit{fully} determine a permutationally invariant state is given by $\binom{d\times d+N-1}{N}$ $-1$, which is approximately $\mathcal{O}(N^{d^2-1})$
~\footnote{The permutationally invariant set of density operators should not be confused with the one formed by convex combinations of permutationally invariant pure states. In the case of $N$ sub-systems of dimension $d$ the set of permutationally invariant pure states form a subspace of dimension $D=$ $\binom{d+N-1}{N}$ \cite{harrow2013church}. Thus, any density operator that is a convex combination of permutationally invariant pure states, will have at most $D\times D-1$ independent parameters. For the case of $N$ qubits, the subspace associated to permutationally invariant pure states will have dimension $N+1$, and then, density operators, which does not necessarily describe a pure state, will have (at most) $(N+1)^{2}-1$ independent (real) parameters.}. Furthermore, given that the MaxEnt estimation technique is designed to work in situations in which an informationally complete data set is not available, in many cases, the number of measurements needed for a reasonable estimation of the density matrix will be even lower than the upper bound mentioned above.
This represents a clear reduction in the resources needed to estimate an unknown symmetric quantum state when compared to the exponential growth of the general case ($\sim\mathcal{O} (d^{2N})$). 
In particular, for three-qubits permutationally invariant states, we have $d=2$ and $N=3$, giving a total of 19 independent real parameters to obtain the density matrix. This is in agreement with the 
number of expectation values of Pauli tensor-product (Fig.~\ref{f:1000QOffOffSIC}(a)) and SIC-POVM (Fig.~\ref{f:1000QOffOffSIC}(b)) observables for which we have obtained a mean fidelity value exactly equal to one. Moreover, with less measurement data, we can still observe a quite good estimation of the state (about 15 observables for a mean fidelity above 0.95).  

It is worth mention that, 
in a quorum situation, i.e., when an informationally complete data set is available, the results presented in this work are comparable with tomographic methods 
designed for permutationally invariant states (see for example Refs.~\cite{PermutationallyInvariantQT,Moroder_2012,Klimov-Permutational-POVM}). 
The use of estimation techniques which are able to provide good results even when the available information is far from quorum, is essential in quantum information theory. This is so, not only because the number of independent measurements needed for a complete tomography grows exponentially with the number of qubits (rendering it impractical in many situations), but also because in some circumstances, it might not be possible to measure certain observables which are required for a given tomographic scheme. In such situations, in which the experimenter \textit{cannot perform} all the experiments needed for obtaining an informationally complete set of measurements for a given state (or a family of states),
the MaxEnt principle should be considered as a good candidate to obtain the least biased estimation. In this sense, our algorithm becomes relevant allowing to easily combine the prior knowledge about symmetries with the MaxEnt principle.

Furthermore, unlike approaches tailored to particular symmetries or families of states, the algorithm proposed here works for arbitrary symmetries. With this aim, we also analyze the case of Werner states, defined by having the symmetry given in Eq.~(\ref{sym_Werner}). To completely determine a Werner state, one can specify the mean values $\mbox{tr}(\rho V_{\pi})$, where $\{V_{\pi}\}$ is the set of matrices associated to the representation of the permutation group \cite{Eggeling_PhD}. Thus, in order to compute how many independent parameters are associated to an $N$-qubit Werner state, one has to compute the number of linearly independent matrices $V_{\pi}$. It is easy to check that for the case of three-qubit states, the set of constraints in Eq.~\eqref{e:ConditionsB} reduces the estimation problem to a $5$-dimensional variety \cite{Eggeling2001}, which means a substantial reduction in the number of independent measurements to be performed with respect to a full tomography. As shown in Figs. \ref{f:1000QOffOffSIC}(c) and \ref{f:1000QOffOffSIC}(d), where the prior information about the symmetries of the state is incorporated to the MaxEnt method, only five expectation values are needed to achieve the maximum fidelity ($F=1$), in agreement with the previous analysis. Moreover, for most states, less than five expectation values are enough in order to obtain a fidelity above $0.95$. 

It is important to 
note that, beyond the global performance of the method, the fidelity of reconstruction depends critically on the particular state to be estimated. Then, the average performance for a given family of states might be quite different from that observed for particular subfamilies. In what follows, we focus on two subfamilies of permutationally invariant states, the so-called \textit{cat-like states} and Dicke states, both of interest in quantum mechanics and for applications in quantum information processing \cite{Ouyang-2014,Hakoshima2020,GHZa,Xu2014}. 

Cat-like states are defined by the superposition
\begin{eqnarray}
|C_{N,p}\rangle=\sqrt{p}|0\rangle^{\otimes N}+\sqrt{1-p}|1\rangle^{\otimes N},
\end{eqnarray}
with $0\leq p \leq 1$. As particular case, the generalized $N$-qubit Greenberger–Horne–Zeilinger state \cite{greenberger1990bell}, defined as $|\mathrm{GHZ}_N\rangle=2^{-1/2}\left(|0\rangle^{\otimes N}+|1\rangle^{\otimes N}\right)$, is a highly entangled cat-like state corresponding to $p=\frac{1}{2}$. We have performed the reconstruction, by using the MaxEnt algorithm with and without symmetries, of different cat-like states with $p=0.1, 0.2, 0.3, 0.4, 0.5$. 
Besides, this set of values of the parameter $p$ corresponds to cat-like states with different degrees of entanglement.
The obtained results for Pauli tensor product operators and SIC-POVM, are displayed in Figs.~\ref{f:Cat}(a)-\ref{f:Cat}(d). For these particular states, we observe also a considerable reduction in the number of observables needed for a good estimation, that is more significant in the case of Pauli measurement basis.
%
\begin{widetext}
\begin{minipage}{\linewidth}
\begin{figure}[H]
\centering
\includegraphics[width=1\textwidth]{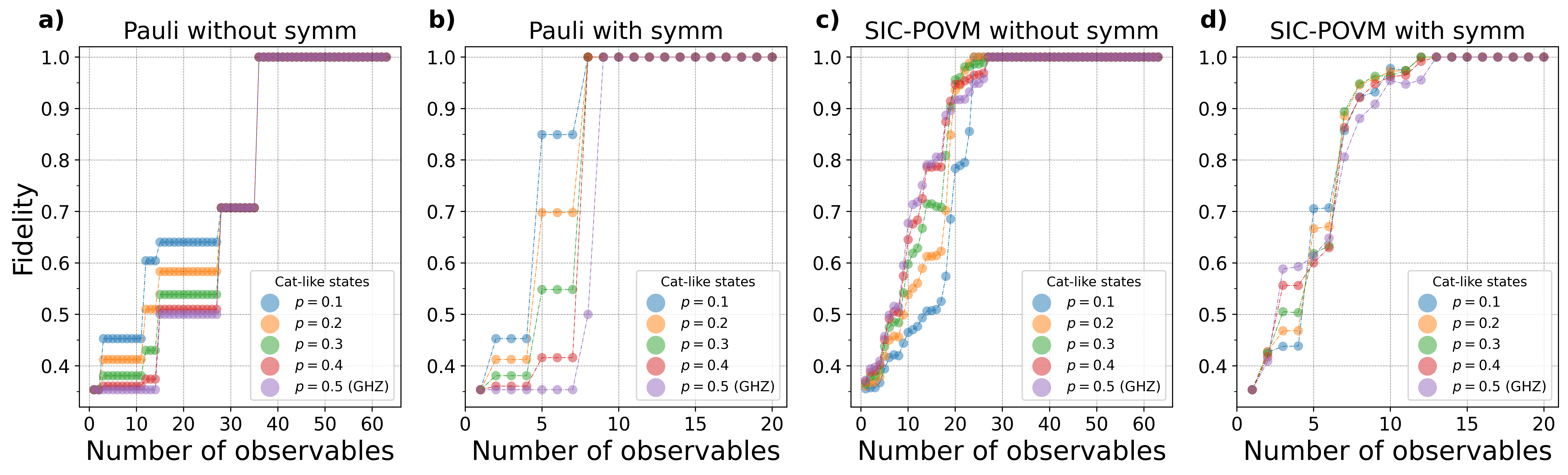}
\vspace{-0.5cm}
\caption{\label{f:Cat}\small{Performance of the quantum MaxEnt estimation for cat-like states of three qubits, $|C_{3,p}\rangle=\sqrt{p}|000\rangle+\sqrt{(1-p)}|111\rangle$. Each point in these figures represents the fidelity of reconstruction obtained from an expectation value set $S_r$, corresponding to a subset of $r$ Pauli (panel \textbf{(a)} and panel \textbf{(b)}) or SIC-POVM (panel \textbf{(c)} and panel \textbf{(d)}) operators.
We followed the standard algorithm (panel \textbf{(a)} and panel \textbf{(c)}) and the algorithm that includes the knowledge of symmetries of the state (panel \textbf{(b)} and panel \textbf{(d)}).  We show the results for five different
weights $p$.}}
\end{figure}
\end{minipage}
\end{widetext}

A great improvement in the estimation was also observed, when including the symmetries, for Dicke states. The $N$-qubit Dicke states~\cite{Dicke1954} with $n$ excitations are defined as
\begin{eqnarray}
|D_{N,n}\rangle= \binom{N}{n}^{-1/2}\sum_k P_k\left(|0\rangle^{\otimes N-n}\otimes|1\rangle^{\otimes n}\right),
\end{eqnarray}
where the summation is over all possible permutations $P_k$. In particular, the Dicke states with $n=1$ corresponds to the generalized W-state, a relevant entangled state which is in a different entanglement class than that of the GHZ-state. The results obtained, with and without symmetries, are displayed in Figs.~\ref{f:Dicke}(a)-\ref{f:Dicke}(d).
%
\begin{widetext}
\begin{minipage}{\linewidth}
\begin{figure}[H]
\centering
\includegraphics[width=1\textwidth]{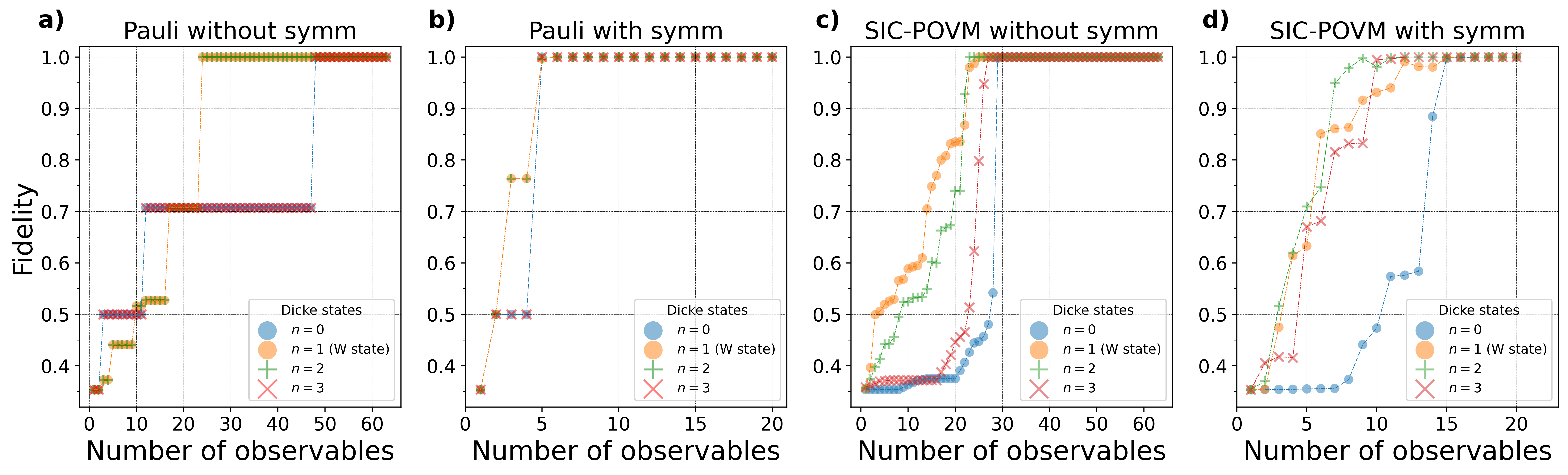}
\vspace{-0.5cm}
\caption{\label{f:Dicke}\small{Performance of the quantum MaxEnt estimation for three-qubit Dicke states, $|D_{3,n}\rangle=\binom{3}{n}^{-1/2}\sum_k P_k\left(|0\rangle^{\otimes 3-n}\otimes|1\rangle^{\otimes n}\right)$. Each point in these figures represents the fidelity of reconstruction obtained from an expectation value set $S_r$, corresponding to a subset of $r$ Pauli (panel \textbf{(a)} and panel \textbf{(b)}) or SIC-POVM (panel \textbf{(c)} and panel \textbf{(d)}) operators.
We followed the standard algorithm (panel \textbf{(a)} and panel \textbf{(c)}) and the algorithm that includes the knowledge of symmetries of the state (panel \textbf{(b)} and panel \textbf{(d)}).  We show the results for four different
Hamming weights $n$.}}
\end{figure}
\end{minipage}
\end{widetext}

All the examples studied in this subsection (Fig.~\ref{f:1000QOffOffSIC}-Fig.~\ref{f:Dicke}) have a particular symmetry. When analyzing the cases of a random sample of Werner or permutationally invariant states (Fig. \ref{f:1000QOffOffSIC}), we find that there are no substantial differences in the smoothness of the fidelity plots for the estimation based on SIC-POVMs compared to that based on Pauli operators. Contrarily, 
in the example of randomly chosen states according to the Haar measure displayed in Fig.~\ref{f:1000MaxEntSolo}, the performance of the method depends substantially on the order of the Pauli operators. A similar behavior is observed for particular subfamilies of permutationally invariant states in Figs. \ref{f:Cat}(a) and \ref{f:Cat}(b), \ref{f:Dicke}(a) and \ref{f:Dicke}(b): there are several observables that do not add substantial information for those states. These examples suggest that, when resorting to the MaxEnt principle to estimate a quantum state being far from the quorum situation,
in some cases, the measurement scheme based on Pauli operators could be optimized by finding an optimum subset of them satisfying that, for a given level of accuracy in the estimation, a minimum number of observables is needed. However, this would require extra computational resources to find, prior to the measurement process or in an adaptive measurement scheme, the optimum set for the particular type of states to be estimated.  On the one hand, we can say that, in general, the performance seems to be less dependent on the ordering of the chosen observables and the type of states to be estimated,
when measurements are based on SIC-POVMs (in the sense that the fidelity curves are smoother). 
On the other hand, given that the number of possible orderings of a list of operators scales as the factorial of its length, the inclusion of symmetries to the estimation of MaxEnt renders the optimization of the set of observables to be used more efficient, since it must be performed on a reduced set of them.

\subsection*{Reliability in the presence of noise}

For the purpose of studying how the MaxEnt method with symmetries will perform in a realistic scenario, we have numerically implemented the estimation process in the presence of different levels of experimental noise, also including the effects of the finite statistics that results from the finite size of the sample. 

As an example of realistic errors in the state preparation, we can consider that the generation process is affected by white noise,
which mixes the multi-qubit pure state $|\psi\rangle$ with the maximally mixed state $I/2^N$:
\begin{eqnarray}\label{eq:rho_noise}
\rho_{noise}\left(|\psi\rangle, \eta\right) = (1-\eta)|\psi\rangle\langle\psi| + \frac{\eta}{2^N}I,
\end{eqnarray}
where the value of the real parameter $\eta$ is a measure of the noise level, related to the purity of the state as $\mathrm{Tr}(\rho_{noise}^2) =\frac{(2^N-1)}{2^N}(\eta-1)^2+\frac{1}{2^N}$. Besides, in the quantum tomographic scenario, measurements are performed on 
$\mathcal{N}$ independently and identically prepared copies of the unknown quantum state. 
%
\begin{widetext}
\begin{minipage}{\linewidth}
\begin{figure}[H]
\includegraphics[width=1\textwidth]{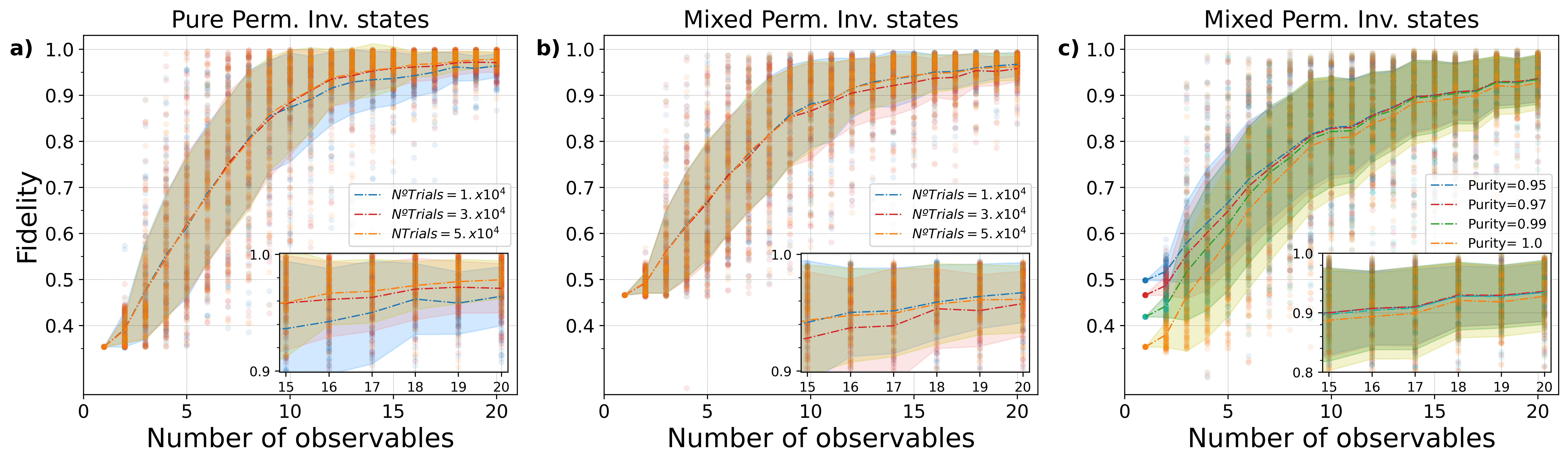}
\vspace{-0.5cm}
\caption{\label{f:Noise}\small{Performance of the MaxEnt estimation method with symmetry constraints in the presence of realistic experimental noise. Each point in these figures represents the value of the fidelity in the reconstruction of a three-qubit permutationally invariant state based on the SIC-POVM. Panel \textbf{(a)} corresponds to an ideal state preparation ($\eta=0$, or equivalently, a purity state equal to 1), a number of dark counts per pulse $\lambda_{dc}=2\times 10^{-4}$, and different number of trials per experiment $\mathcal{N}=1\times10^{4}$ (\textit{blue}), $3\times 10^{4}$(\textit{red}), $5\times 10^{4}$ (\textit{orange}). In panel \textbf{(b)}, we have kept the same number of trials that in panel \textbf{(a)} ($\mathcal{N}=1\times10^{4}$ (\textit{blue}), $3\times 10^{4}$(\textit{red}), $5\times 10^{4}$ (\textit{orange})) but considering a purity state equal to 0.97 and $\lambda_{dc}=5\times 10^{-4}$. In panel \textbf{(c)}, the results were obtained for fix values $\lambda_{dc}=1 \times 10^{-4}$ and $\mathcal{N}=10^4$ while varied the value of the purity state (0.95 (\textit{blue}), 0.97 (\textit{red}), 0.99 (\textit{green}), 1.0 (\textit{orange})).
The algorithm was run for 100 states randomly chosen. The dot-dashed lines represent the mean value of the fidelity and the shaded area indicates its standard deviation.}}
\end{figure}
\end{minipage}
\end{widetext}
Then, for each observable $A_i$, the expectation value of the target state, $a_i^{\rho}=\mathrm{Tr}(\rho A_i)$, is approximated by the relative frequency $f_i = n_i/\mathcal{N}$, experimentally obtained, where $n_i$ is the number of trials that resulted in a click of the detectors. In agreement with typical experimental setups based on optical platforms \cite{Goyeneche2015,Stefano:19},
we have assumed a source consisting of a
pulsed attenuated laser, that emits weak coherent states at the single-photon level. 
In this context, it is reasonable to assume that the statistics of the photon emission process and the dark counts, caused by self triggering effects in the photon detectors, are Poissonian.
In such a case, the probability of detecting a photon in the mode $|\psi_k\rangle$ is given by
\begin{eqnarray}
\mathcal{P}_k(1\,\mbox{count})=1-\exp^{-\mu p_k-\lambda_{dc}},
\end{eqnarray}
where $p_k=\mathrm{Tr}(\rho_{noise} |\psi_k\rangle\langle\psi_k|)$ is the corresponding ideal probability, while $\mu$ and $\lambda_{dc}$ are the mean number of photons and dark counts, per pulse, respectively.
For a fix number of trials (or pulses) $\mathcal{N}$, the detected counts follow a binomial distribution
governed by a mean number of counts $\bar{n}_k=\mathcal{N}( 1-\exp^{-\mu p_k-\lambda_{dc}})$. 

By incorporating all those features in our codes, we have obtained the results displayed in Fig.~\ref{f:Noise}. The numerical simulations where performed for a value $\mu= 0.18$ that results in 84$\%$ of empty pulses and
about 2$\%$ of pulses with more than one photon.
In all cases we have first generated 100 
pure states randomly chosen from a sample of permutationally invariant states, as our desired ideal state $|\psi\rangle$. From these, noisy states with different degree of purity were prepared according to Eq.~(\ref{eq:rho_noise}). The estimation process was run taking into account the symmetry of the states and the expectation values of a number of $r$ SIC-POVM operators. Then, the fidelity $F$ between $\rho_r^{\mathrm{SME}}$ and $\rho_{noise}$ is calculated. In Fig.~\ref{f:Noise}(a) we can see the effect of the finite statistic on the estimation of pure states ($\eta=0$), when the number of dark counts per pulse is $\lambda_{dc}=2\times 10^{-4}$. This last value is consistent with typical experimental situations, i.e., a pulse duration in the order of microseconds and a single photon detector module operating in the range of 100 dc/sec. In Fig.~\ref{f:Noise}(b) we show the estimated states for the same statistics that in Fig.~\ref{f:Noise}(a) but for a higher noise level both in the preparation stage, where now the initial states have a purity of $0.97$, and in the detection stage where $\lambda_{dc}=5\times10^{-4}$. Finally, in Fig.~\ref{f:Noise}(c), it is shown how the quality of the estimation is affected by a noisy preparation of the states for a given number of trials ($\mathcal{N}=1\times10^4$) and dark counts ($\lambda_{dc}=1\times10^{-4}$). It can be seen that the mean fidelity increases for higher level of noise in the preparation stage, i.e., when the purity of the states is reduced. This is consistent with the principle in which is based our estimation method: as the simulations were carried out with the addition of white noise to the pure target state, the prepared state will not be pure and therefore its entropy will increase, but also it will be the one that maximizes the entropy for a given level of noise, since that noise is introduced without any bias. Hence, the MaxEnt estimator performs better when such a model of noise (white noise) begins to be dominant over other types of noise models.

According to these results we conclude that, while for some states the performance of the MaxEnt estimation under symmetry constraints is clearly affected by noise, the mean value of the fidelity behaves in a quite robust way, and a similar behavior is observed for the standard deviation. This means that, with a high probability, the method would allow to obtain a reasonable fidelity between the prepared and the estimated state, in realistic scenarios that are relevant for quantum information processing tasks.

\section{Conclusions}\label{conclusions}

In this work we have presented an algorithm which performs the quantum MaxEnt estimation incorporating symmetry constrains. The method works for arbitrary symmetries of the state to be estimated, having as inputs the generators of the symmetry under consideration, without any additional dependence on other mathematical features of the underlying group. We have illustrated how it works and studied its performance in different examples of interest for quantum information tasks. The fact that the method is presented in an algorithmic way allows us to carry out a systematic study for different families of states and measurement bases. We carried out numerical experiments to validate the technique and showed that the estimator is robust against typical experimental noises, such as those present in the generation of the states, those due to the imperfections of the measurement devices, and those originated in the statistics obtained when dealing with finite samples.

Although we restricted our attention to multi-qubit systems and showed the results for the case of three qubits, the method is completely general and can be used to estimate the state of higher-dimensional systems. 
Our work generalizes previous tomographic methods specially developed to the reconstruction of quantum states with a particular symmetry. In general, those methods allow to reduce the number of measurements required with respect to a standard QST. But even so, they are based on a set of informationaly complete measurements related to the subspace with the given symmetry. 
In our case, the estimation process consists in an inference method based on the MaxEnt principle. Therefore, even when we do not have access to all the measurements for a complete reconstruction,  
we can still find an estimation (the most reliable one) of the unknown state.
As our results clearly show, the inclusion of the prior information encoded in the symmetries could significantly improve such estimation for a given number of observables measured.

 While it is usually reported that the MaxEnt method has a good performance for quantum state estimation, our results reveal that the notion of ``good performance" requires further specification. In fact, in scenarios where the source generates states with a given symmetry and this is not taken into account, the performance of the standard MaxEnt method drops off in comparison to the case where the source generates states which are uniformly distributed over the Hilbert space (unbiased source). This important point was not discussed in previous publications. In addition, we have shown that the performance strongly depends on the set of observables to be measured and the particular state to be estimated. This opens the problem of optimizing (or adaptatively acquired) the measurement set based on previous knowledge about the source, as can be done, for example, following the procedures presented in Refs.~\cite{qi2017adaptive,cao2020neural}. Even so, by including the symmetry information of the state in the MaxEnt method, the number of measurements needed for a good estimation is considerably less than that needed for a standard QST, having as an upper bound the number needed in other methods specially developed for a particular symmetry (as is the case of permutationally invariant states). However, finding the estimated state becomes computationally hard as the dimension of the system grows, due to the nonlinear optimization problem associated to the MaxEnt method. For this reason, faster techniques as variational methods have been considered~\cite{maciel2011variational}. In any case, the study of the MaxEnt method is important in itself, because it operates at the heart of statistical inference: it provides the least biased option compatible with the available information. Also, it works as the reference for other reconstruction techniques with incomplete data, since these are proposed under the assumption that they produce similar results to those of MaxEnt~\cite{Goncalves-2013-MaxEntTomography}.

Finally, it is important to remark that the MaxEnt method with symmetry constraints could be combined with other methods for quantum state estimation
For example, it is natural to use the MaxEnt method in connection with the maximum likelihood estimation in order to systematically select the most likely estimator with the largest entropy~\cite{Teo-2011}, a subject that we will address in a future work~\cite{ChinosFuture}. 

\label{key}

\begin{acknowledgments}
We express our gratitude to Dr. Q. P. Stefano for helpful discussions about experimental implementations and noisy models.
We also thank Prof. R. Giuntini and Prof. G. Sergioli for providing us with the computer lab server of the Quantum Cagliari group.   
F.H. was partially funded by the project “Per un’estensione semantica della Logica Computazionale Quantistica- Impatto teorico e ricadute implementative”, Regione Autonoma della Sardegna, (RAS: RASSR40341), L.R. 7/2017, annualità 2017- Fondo di Sviluppo e Coesione (FSC) 2014–2020. 
\end{acknowledgments}

\bibliography{main}

\end{document}